\documentclass[12pt]{article}
\usepackage{psfig}

\begin{document}
\sloppy

\title{The Lamellar-to-Isotropic Transition in Ternary 
Amphiphilic Systems}          
\author{U. S. Schwarz$^{(a)}$, K. Swamy$^{(b)}$ and G. Gompper$^{(a,b)}$\\
$^{(a)}$ Max-Planck-Institut f\"ur Kolloid- und 
                               Grenzfl\"achenforschung \\
Kantstr. 55, 14513 Teltow, Germany \\
$^{(b)}$ Sektion Physik der Ludwig-Maximilians-Universit\"at 
                            M\"unchen \\
Theresienstr. 37, 80333 M\"unchen, Germany } 

\date{}
\maketitle

\begin{abstract}
We study the dependence of the phase behavior of ternary amphiphilic
systems on composition and temperature.  Our analysis is based on a
curvature elastic model of the amphiphile film with sufficiently large
spontaneous curvature $c_0$ and sufficiently negative saddle-splay
modulus $\bar\kappa$ that the stable phases are the lamellar phase
and a droplet microemulsion.  In addition to the curvature energy, we
consider the contributions to the free energy of the long-ranged 
van der Waals interaction and of the undulation modes.  We find
that for bending rigidities of order $k_BT$, the lamellar phase
extends further and further into the water apex of the phase diagram
as the phase inversion temperature is approached, in good agreement
with experimental results.
\end{abstract}

\medskip
\begin{center}
\end{center}

\newpage

\section{Introduction}

\par The theory of ternary amphiphilic systems has developed rapidly
during the last few years \cite{gg:gomp94d,gelb95}. For long-chain 
amphiphiles the molecular solubility is very small, so that essentially
all amphiphilic molecules are located in monolayers at the interface
between oil and water. In this long-chain limit, 
the description of the system by an ensemble of surfaces is justified and 
used frequently \cite{Safran}. The monolayer is described by an ideal 
surface in this approach, with the shape and fluctuations determined by
the curvature energy \cite{helf73}
\begin{equation}
{\cal H}_{curv} = \int dS \left[ 2\kappa (H-c_0)^2 + \bar\kappa K \right]
\end{equation}
where $H$ and $K$ are the mean and Gaussian curvatures, respectively,
and the integral runs over the total interface area. The parameters 
are the bending rigidity $\kappa$, the saddle-splay modulus $\bar\kappa$,
and the spontaneous curvature $c_0$.

\par Most of the previous theoretical work on the phase behavior of
ternary amphiphilic systems has focused on balanced systems with
$c_0=0$, and on the interplay between the curvature energy, which
favors the lamellar phase, and entropy, which favors disordered
bicontinuous or droplet phases \cite{gg:gomp94d,Safran}.
However, in addition to the curvature energy, long-range interactions
between the molecules also give an important contribution to the free
energy.  This is well known for the unbinding of membranes
\cite{lipo95}, and for wetting transitions in simple fluids
\cite{diet88}.  Attractive interactions have also been used to 
explain the reentrant phase separation of spherical and
(very long and flexible) cylindrical micelles near the water apex of
the composition phase diagram \cite{mene95}.  In systems containing
non-ionic amphiphiles, or in systems with screened electrostatic
interactions, the dominant contribution is the van der Waals
interaction \cite{maha76,isra92}.

\par In this paper, we want to study the dependence of phase diagrams
on the spontaneous curvature $c_0$.  We consider systems with $c_0$
sufficiently large that the isotropic phase is a droplet
microemulsion (in order to avoid complications with the bicontinuous
phase).  It has been shown experimentally that $c_0$ varies linearly
with temperature $T$ over a wide range of $T$
\cite{stre94}.  Using this relation for $c_0(T)$, we can
compare our results directly with the temperature dependence of
experimental phase diagrams.

\par It has been shown \cite{safr84} that in the zero-temperature
limit, the curvature model --- together with the constraints of fixed
volume fractions $\phi_s$ of the amphiphile and $\phi_o$ of the
interior phase --- predicts for $-0.5 < \bar\kappa /\kappa <0$ with
increasing $c_0 \phi_o/\phi_s$ the progression from a lamellar phase
via a phase of infinitely long cylindrical micelles to a phase of
spherical micelles, while for $ \bar\kappa /\kappa < -0.5 $ the phase
of cylindrical micelles is absent.  This model has been extended to
include both thermal fluctuations and long-range interactions in
Ref.~\cite{roux86}.  However, with the interactions described
phenomenologically, and no explicit temperature dependence,
this model yields highly schematic phase diagrams, which cannot be
compared easily with experiment.

\section{The Model}

\par We consider the coexistence of a lamellar phase $L_\alpha$ and an
oil-in-water microemulsion phase $L_1$.  We assume that $\bar\kappa$
is sufficiently negative that a hexagonal phase of cylindrical
micelles is suppressed \cite{safr84}.  The $L_\alpha$-phase is
a stack of oil- and water layers of thickness $d$ and $l$,
respectively.  All lengths are measured in units of the amphiphile
length, so that the repeat distance in the lamellar phase is given by
$L=d+l+2$.  The radius of the neutral surface of the oil droplets is
denoted $R$, so that the radius of the oil core is $R-1/2$, while the
outer radius of the micelle is $R+1/2$ (for equal lengths of heads and
tails).  The number density of droplets is $\rho$.

\par We write the free-energy density, $f$, of the lamellar and the 
droplet phases as a sum of three contributions,
\begin{equation} \label{superposition}
f = f_{curv} + f_{therm} + f_{vdW}.
\end{equation}
The curvature-energy density, $f_{curv}$, is given by 
\begin{equation}
f_{curv} = \left\{ \begin{array}{ll}
4 \kappa c_0^2 / L & \mbox{for $L_\alpha$} \\
4\pi \rho \left(2\kappa (1-c_0 R)^2 + \bar\kappa \right) & 
\mbox{for $L_1$}
\end{array} \right.
\end{equation}
Thermal fluctuations in the lamellar phase lead to a steric
(repulsive) interaction, as first proposed by Helfrich \cite{helf78}.
When the size of the amphiphile is used as an ultraviolet cutoff for
the wavelengths of the undulation modes, the steric-interaction 
density of pairs of monolayers, which are separated by a water film 
of thickness $l$, is found to be
\begin{equation}
  \label{HelfrichWW_mod}
 f_{steric,w}(l)  = \frac{k_B T}{8 \pi L} 
      \left( q_c^2 \ln [1 + (q_c \xi_{\parallel})^{-4}]  + 
 \frac{2}{\xi_{\parallel}^2} \arctan [(q_c \xi_{\parallel})^2] \right) \ ,
\end{equation}
where $q_c=2\pi$, and 
$\xi_{\parallel} = (8 \kappa \mu / k_B T)^{1/2} \ l$ is the parallel
correlation length.  The parameter $\mu$ has been determined by Monte
Carlo simulations to be $\mu \simeq 0.135$ \cite{gg:gomp89b,netz95a}.
Note that Eq.~(\ref{HelfrichWW_mod}) reduces to the familiar
$l^{-2}$-form of the steric interaction of {\it two} monolayers
\cite{helf78} for $q_c \xi_\parallel \gg 1$. 

\par It has been pointed out recently that for an {\it asymmetric} 
lamellar
phase, the steric interaction of two monolayers is a function of {\it
both} the thickness of the oil and of the water films \cite{netz95b}.
This is obvious in the case of vanishing oil concentration, where the
interacting surfaces are amphiphilic bilayers with an effective
bending rigidity $\kappa_{eff} = 2\kappa$.  In Ref.~\cite{netz95b},
Monte Carlo simulations have been used to extract the expression
\begin{equation}
  \label{Netz}
  \kappa_{eff}(l;d)/\kappa = 2-(d/l)^\nu / 
               \left[ (d/l)^\nu + {x_0}^\nu \right] 
\end{equation}
with $\nu = 3/2$ and $x_0 = 0.3$. The steric interaction, 
$f_{steric,o}(d)$, of two monolayers through an oil film is obtained 
from Eqs.~(\ref{HelfrichWW_mod}) and (\ref{Netz}) by interchanging
$l$ and $d$. Thus, the total entropic part of the free-energy density 
of the lamellar phase is
$f_{therm}(l,d) = f_{steric,w}(l) + f_{steric,o}(d)$.

\par To calculate the entropy of the droplet phase, we assume that the 
swollen
micelles behave as hard spheres of radius $R+1/2$. The free-energy 
density is then very well approximated by the Carnahan-Starling form 
\cite{hans86}
\begin{equation}
\label{drop_steric}
  f_{HS}(R,\rho) = k_B T \rho 
   \left( \ln \Lambda^3 \rho -1 + \frac{4-3\eta}{(1-\eta)^2} \eta \right)
\end{equation}
where $\eta = (4 \pi / 3) \rho (R+1/2)^3$ is the packing fraction, and 
$\Lambda$ a microscopic length scale, which in the following is taken to 
be equal to the amphiphile length. Undulation modes on 
microemulsion droplets give an additional contribution to the free energy, 
which in the limit $R \ll c_0^{-1}$ has the form \cite{mors94a}
\begin{equation} 
f_{drop}(R,\rho) = \rho \frac{8}{3} k_B T \ln(R^2) \ .
\end{equation}
Such a term becomes important for large droplet radii, and disfavors
the droplet microemulsion in this regime. The entropic part of the 
free-energy density of droplets then reads
$f_{therm}(R,\rho) = f_{HS}(R,\rho) + f_{drop}(R,\rho)$.

\par Finally, we have to calculate the contribution of the
van der Waals interaction to the free energy.  This is done within the
Hamaker approach, where a pairwise form of the interaction is assumed
\cite{maha76,isra92}.  The dielectric properties of the amphiphile 
tails (heads)
are assumed to be identical to those of the oil (water) molecules.  We
subtract the free energy of the pure water phase, so that only the
interaction between oil molecules (and amphiphile tails) has to be
taken into account.  For a molecular fluid, the interaction 
free-energy density is given by \cite{hans86}
\begin{equation} \label{vdW_mol}
f_{vdW} = \frac{1}{V} \int_V d^3 r_1 \int_V d^3 r_2 \ 
                        w(|{\bf r_1-r_2}|) g(|{\bf r_1-r_2}|)
\end{equation}
where $w(r)$ is the interaction potential, which consists of a
long-ranged attractive and a short-ranged repulsive part, and $g(r)$
is the normalized density-density correlation function, which
approaches unity for large distances and vanishes for short ones.  We
approximate the product of these two functions by
\begin{equation} \label{interact}
w(r) g(r) = \phi(r) \equiv \frac{A}{\pi^2} \left[
 \frac{\epsilon^2}{(r^2 + \epsilon^2)^4} - \frac{1}{(r^2+\epsilon^2)^3} 
 \right]
\end{equation}
where $A$ is the Hamaker constant.
The parameter $\epsilon$ in Eq.~(\ref{interact}) determines at which 
length scale (in units of the amphiphile length) the repulsion becomes 
dominant. 

\par The calculation of the free-energy density, $f_{intra}$, of the
van der Waals interaction of the oil molecules within one aggregate is
straightforward in both the droplet and lamellar phases (compare
Ref.~\cite{maha76}).  This also holds for the inter-aggregate part of
the free-energy density, $f_{inter}$, in the lamellar phase.  In the
(disordered) droplet phase, the inter-aggregate part is taken to be
\begin{equation} \label{inter_drop}
f_{inter,drop} = \frac{\rho^2}{2} \int_{r>2R+1} d^3r \ W_{R}(r)
\end{equation}
where $W_{R}(r)$ is the interaction energy of two spheres of radius
$R$ with distance $r$ between their centers.  To obtain
Eq.~(\ref{inter_drop}), we have replaced the two-particle distribution 
function for hard spheres of radius $R+1/2$ by a step function
\cite{cous95}.  It is important to note that with the choice
(\ref{interact}) for the interaction potential, all interaction
integrals can be carried out analytically.  However, the results are
rather lengthy and can therefore not be given here.

\section{Phase Diagrams}

\par In order to calculate phase diagrams, we add chemical potential
terms to the free energy, and minimize with respect to $l$ and $d$, or
$R$ and $\rho$, respectively.  Since the functional form of
$\kappa_{eff}(l;d)$ was determined self-consistently in
Ref.~\cite{netz95b}, we have to use the same procedure here.  The free
energies in our model depend on the parameters $\kappa$, $\bar\kappa$,
$c_0$, and $A$.  The value of the Hamaker constant is well known
experimentally to be $A = (0.28 T/300K + 0.17) 10^{-20} J$ for the
interaction of hydrocarbon across water \cite{isra92}.  For $T=300 K$
one therefore has $A\simeq 1k_BT$.  For the spontaneous curvature, we
assume a linear temperature dependence of the form $c_0(T) = c_0(T_1)
(T_2 - T)/(T_2-T_1)$, where $T_2$ denotes the phase-inversion
temperature.

\par A typical sequence of phase diagrams with decreasing spontaneous
curvature $c_0$ is shown in Fig.~1 for bending rigidities of the order
of $1 k_B T$.  The parameters are chosen such as to allow a comparison
with the experimentally well studied oil-water-$C_{12}E_5$ system.  In
addition to the two-phase coexistence region of lamellar and droplet
phase, we also show the two-phase coexistence of the droplet
microemulsion with a pure oil-phase (emulsification failure).  We can
follow the two-phase coexistence curves in the $L_1$-phase only to a
minimum water concentration of about 30\%, where the oil-filled
micelles become close-packed.  The most prominent feature of the phase
diagrams shown in Fig.~1 is that with decreasing spontaneous
curvature, the lamellar phase extends further and further into the
water apex of the Gibbs triangle.  This behavior is in excellent
agreement with experimental observations \cite{kuni82a,olss93}.

\par Phase diagrams for larger bending rigidities are shown in Fig.~2.
For $\kappa/k_BT \simeq 3$, the reentrant behavior of the droplet phase
disappears, and the lamellar phase is found to be always stable for 
small oil content. Finally, for $\kappa/k_BT \simeq 10$, the lamellar
phase (with low oil content) can no longer be swollen to large 
intermembrane separations, but coexists with an almost pure water phase.

\section{Discussion}

\par The extension of the lamellar phase into the water apex of the
Gibbs phase triangle for $\kappa/k_BT \simeq 1$ can be understood as
follows.  Consider a path in the Gibbs triangle of Fig.~1b of constant
amphiphile concentration, for which the system is in the droplet phase
for equal oil- and water concentrations.  With decreasing oil content,
the droplet size decreases, so that the bending energy cost for
droplets increases, and ultimately leads to a transition into the
lamellar phase.  This is the effect considered in Ref.~\cite{safr84}.
With further decreasing oil content, the steric interaction of the
monolayers across the oil films increases, and leads to a (reentrant)
transition to the droplet phase.

\par When the bending rigidity is increased to $\kappa/k_BT \simeq 3$,
the lamellar phase can be swollen by a very large amount by adding
water.  Thus, the repulsive interaction dominates in the lamellar
phase for $\kappa/k_BT {< \atop \sim} 3$.  Only when the bending
rigidity is increased to $\kappa/k_BT > 3$ does the attractive
interaction begin to affect the phase behavior, and unbinding
transitions \cite{lipo95} can be observed (compare
Refs.~\cite{leib87b} for a discussion of unbinding transitions in the
binary water-amphiphile system).  The first-order character of the
phase transitions for $\kappa/k_BT {< \atop \sim} 3$ discussed above
should therefore hold beyond the superposition approach
(\ref{superposition}).

\par It is also interesting to note that the
van der Waals interaction is {\it not} strong enough to produce a
two-phase coexistence between two droplet microemulsion, in agreement
with recent calculations for colloidal systems \cite{cous95}.  This
result may change when the deformability of the oil droplets is taken
into account, which allows a larger contact area of two adhering
droplets and thereby increases the interaction energy \cite{dano93}.

\par Our work can be extended in several directions; it would be
interesting, for example, to investigate the stability of the hexagonal
phase of cylindrical micelles.  A more detailed comparison with
experiment requires the systematic measurement of ternary phase
diagrams with larger bending rigidities.

\bigskip  
\noindent {\bf Acknowledgments:}
\smallskip
\noindent
We thank J. Goos, H. L\"owen and R. Strey for stimulating discussions.

\begin{figure}[htbp]
  \begin{center}
    \leavevmode
    \psfig{file=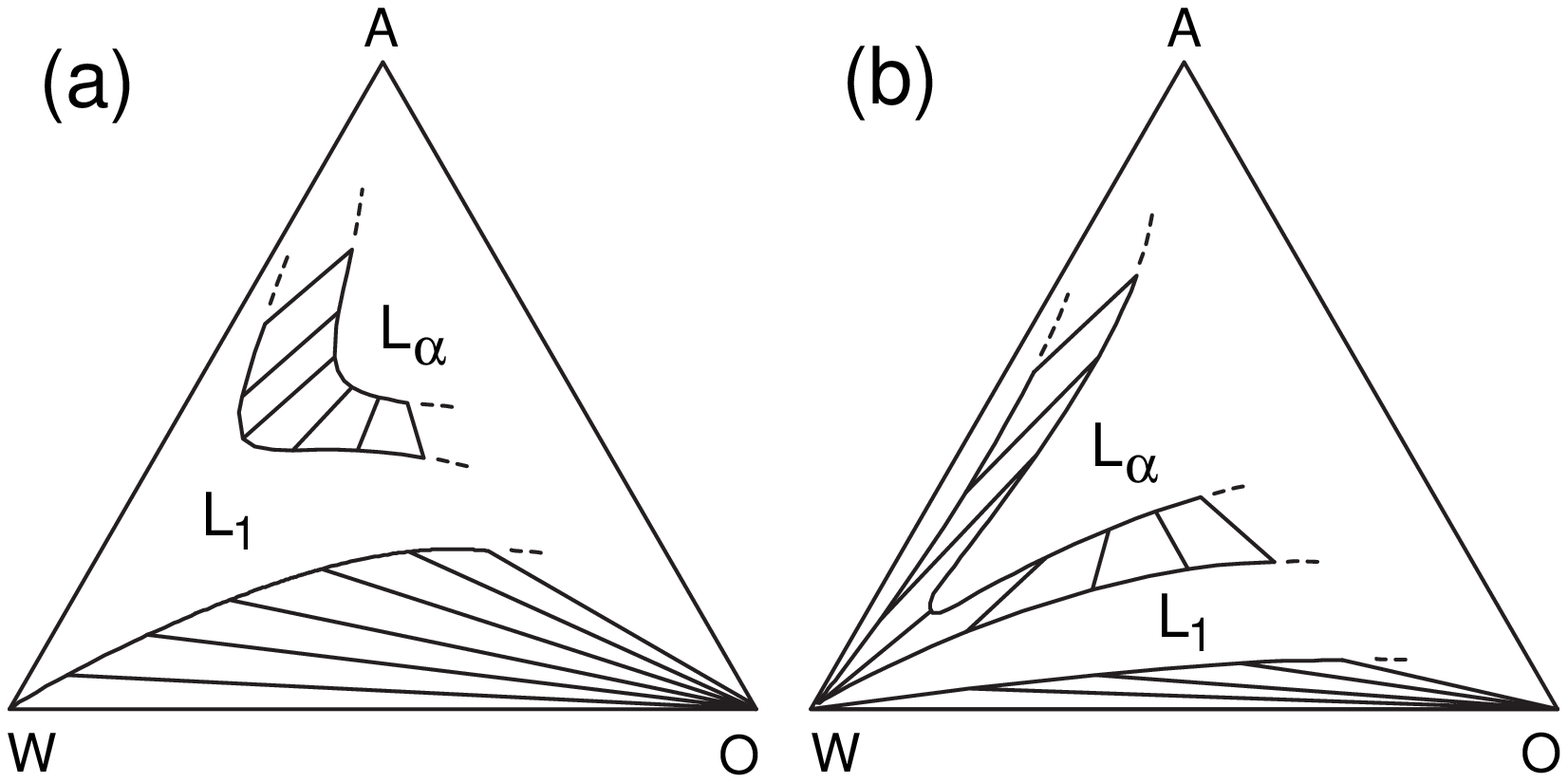}
  \end{center}
  \caption{Ternary composition phase diagrams for low bending
rigidity and (a) $T = 293 K$ and (b) $T = 310 K$.  The parameters of
the model are $\kappa/k_BT_1=1.0$, $\bar{\kappa}/\kappa=-0.5$,
$c_0(T_1)=1/6$, $\epsilon = 0.7$, $T_1 = 293 K$ and $T_2 = 321 K$.
Each Gibbs triangle contains two coexistence regions:  for higher
amphiphile concentrations the lamellar phase $L_\alpha$ coexists with
the droplet phase $L_1$, for lower amphiphile concentration the
droplet phase coexists with an excess oil phase (emulsification
failure).}
  \label{fig:bild1}
\end{figure}

\begin{figure}[htbp]
  \begin{center}
    \leavevmode
    \psfig{file=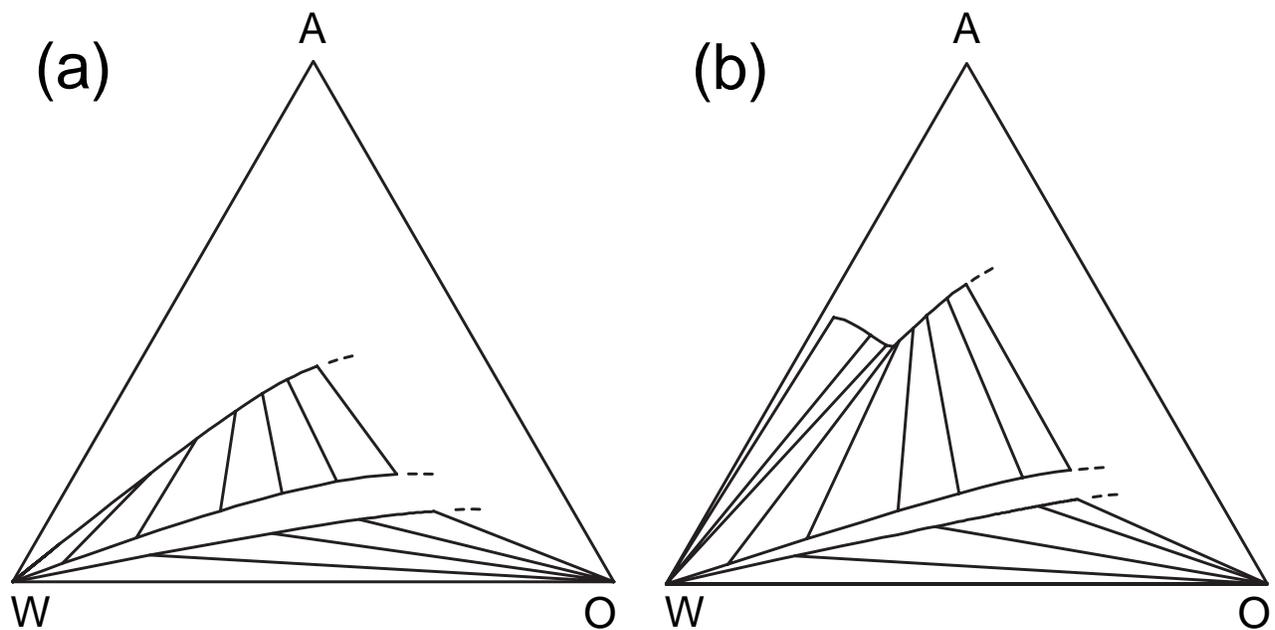}
  \end{center}
  \caption{Ternary composition phase diagrams for (a) intermediate
and (b) high bending rigidity.  The parameters are $c_0^{-1}=15.3$,
$\bar{\kappa}/\kappa=-0.5$, $\epsilon = 0.7$, and $\kappa/k_BT=2.8$
and $\kappa/k_BT=9.5$, respectively.}
  \label{fig:bild2}
\end{figure}

\end{document}